*Article*

# Two Strategies for Boreal Forestry with Goodwill in Capitalization

**Petri P. Kärenlampi** [1*]

[1] Lehtoi Research, Finland
\* Correspondence: petri.karenlampi@professori.fi

**Abstract:** Two strategies for boreal forestry with goodwill in estate capitalization are introduced. A strategy focusing on Real Estate (RE) is financially superior to Timber Sales (TS). The feasibility of the RE requires the presence of forest land end users in the real estate market, like insurance companies or investment trusts, and the periodic boundary condition does not apply. Commercial thinnings do not enter the RE strategy in a stand-level discussion. However, they may appear in estates with a variable age structure and enable an extension of stand rotation times.

## 1. Introduction

For decades, forest estates have been lucrative investments [1,2,3,4,5,6,7,8,9]. Timber sales proceedings have developed conservatively [10,11,12,13], but there has been a significant development in the valuation of estates [2,5,8,9]. The development of estate valuation probably has been related to declining market interest rates, impairing yields from interest-bearing instruments [14,15]. It is thus suspected that the inflated capitalizations are due to factors external to the forestry business [cf. 16,17,18,19,20]. Also, vertical integration within the forestry sector may induce a valuation premium for forest estates [21,2]. In addition, private-equity timberland often appears as a favorable component in diversified portfolios [22,2,23,24].

An ownership change has been related to the increased estate valuation. In North America and in the Nordic Countries, institutions concentrating on the business of investing have purchased timberland from forest products companies [25,21,1,2]. Forestry institutions, rather than private individuals, have recently dominated the estate market [26,9,7]. Some investment firms have included carbon sequestration in their business strategies [27,28]. However, enhanced carbon sequestration generally induces a deficiency in the gained financial benefit [29,30,31,32,33].

Forestry investments have been recently analyzed in terms of financial economics. However, private-equity timberland returns are poorly explained by Capital Asset Pricing Model CAPM [25,2], even if stumpage prices appear to support timberland returns [34]. Improving investor sentiment impairs timberland returns [35]. Arbitrage Pricing Theory (APT) is a complicated approach, including an intuitive selection of explaining factors [25,2].

The increased capitalization contributes to the financial return in operative forestry. Greater valuation necessarily reduces the return of capital invested. The greater valuations may contribute to the feasible management practices. Change in valuations also may contribute to the financial deficit induced by enhanced carbon sequestration, biodiversity advancement, or recreational modifications.

Instead of merely referring to average market prices of forest estates, valuations in terms of tangible and intangible value components appearing on forest stands and estates have been recently discussed: trees, land, amortized investments, and eventual goodwill values [36]. Observations have indicated that the proportional goodwill is close to reality within the Nordic Region, however resulting in continuity problems [36]. There is some theoretical justification for the appearance of the premium [36]. It has been found that goodwill value deteriorates along with harvesting. Such deterioration however can be at least partially avoided by exploiting the real estate market, instead of merely the timber market [36].

In the remaining part of this paper, we will first review the financial theory, including the consequences of inflated capitalization. A hybrid Equation is developed, allowing but not forcing commercial thinnings. Two strategies are introduced, one focusing on entering the Real Estate market (RE), and the other on Timber Sales (TS), the former applying the hybrid Equation. Then, experimental materials are described. Financial analysis is implemented for the performance of the two strategies, and results are discussed.

## 2. Materials and Methods

### 2.1. Financial considerations

We apply a procedure first mentioned in the literature in 1967, but applied only recently [37,38,39,40,41,42,43,32,33,36]. Instead of discounting revenues, the capital return rate achieved as relative value increment at different stages of forest stand development is weighed by current capitalization, and integrated.

The capital return rate is the relative time change rate of value. We choose to write

$$r(t) = \frac{d\kappa}{K(t)dt} \quad (1)$$

where $\kappa$ in the numerator considers value growth, operative expenses, interests, and amortizations, but neglects investments and withdrawals. In other words, it is the change of capitalization on an economic profit/loss basis. $K$ in the denominator gives capitalization on a balance sheet basis, being directly affected by any investment or withdrawal. Technically, $K$ in the denominator is the sum of assets bound on the property: bare land value, the value of trees, and the non-amortized value of investments. In addition, intangible assets may appear. The pricing of forest estates may include goodwill value.

The momentary definition appearing in Eq. (1) provides a highly simplified description of the capital return rate. In reality, there is variability due to several factors. Enterprises often contain businesses distributed to a variety of production lines, geographic areas, and markets. In addition, quantities appearing in Eq. (1) and are not necessarily completely known but may contain probabilistic scatter. Correspondingly, the expected value of capital return rate and valuation can be written, by definition,

$$\langle r(t)\rangle = \frac{\int p_{\frac{d\kappa}{dt}} \frac{d\kappa}{dt} d\frac{d\kappa}{dt}}{\int p_K K(t) dK} = \frac{\int p_{\frac{d\kappa}{dt}} r(t) K(t) d\frac{d\kappa}{dt}}{\int p_K K(t) dK} \quad (2)$$

where $p_i$ corresponds to the probability density of quantity $i$.

Let us then discuss, the determination of capital return rate in the case of a real estate firm benefiting from the growth of multiannual plant stands of varying ages. Conducting a change of variables in Eq. (3) results as

$$\langle r(t)\rangle = \frac{\int p_a(t) \frac{d\kappa}{dt}(a,t) da}{\int p_a(t) K(a,t) da} = \frac{\int p_a(t) r(a,t) K(a,t) da}{\int p_a(t) K(a,t) da} \quad (3).$$

where $a$ refers to stand age. Eq. (3) is a significant simplification of Eq. (2) since all probability densities now discuss the variability of stand age. However, even Eq. (3) can be simplified further.

In Eq. (3), the probability density of stand age is a function of time, and correspondingly the capital return rate, as well as the estate value, evolve in time. A significant simplification would occur if the quantities appearing on the right-hand side of Eqs. (2) and (3) would not depend on time. Within forestry, such a situation would be denoted "normal forest principle", corresponding to evenly distributed stand age determining relevant stand properties [44].

$$\langle r(t)\rangle = \frac{\int \frac{d\kappa}{dt}(a) da}{\int K(a) da} = \frac{\int r(a) K(a) da}{\int K(a) da} \quad (4).$$

The "normal forest principle" is rather useful when considering silvicultural practices, but seldom applies to the valuation of real-life real estate firms, with generally non-uniform stand age distribution. However, it has recently been shown [32] that the principle is not necessary for the simplification of Eq. (3) into (4). This happens by focusing on a single stand, instead of an entire estate or enterprise, and considering that time proceeds linearly. Then, the probability density function $p(a)$ is constant within an interval $[0,\tau]$. Correspondingly, it has vanished from Eq. (4).

Application of Eqs. (1) to (4) does require knowledge of an amortization schedule. Here, regeneration expenses are capitalized at the time of regeneration and amortized at the end of any rotation [43].

By definition, inflation of capitalization corresponds to the emergence of a surplus in the capitalization $K$ appearing in the denominator of Eqs. (1) to (4). Simultaneously, the value change rate $d\kappa/dt$ in the numerator may or may not become affected.

Before discussing the details of inflated capitalization, a periodic boundary condition is given as

$$\int_a^{a+\tau} \frac{dK}{dt} dt = 0 \quad (5),$$

where $\tau$ is rotation age. On the other hand, the value growth rate sums up as free cash flow as

$$\int_{a}^{a+\tau} \frac{d\kappa}{dt} dt = \int_{a}^{a+\tau} \frac{dC}{dt} dt \qquad (6),$$

where $dC/dt$ refers to the rate of free cash flow from timber sales.

Let us then discuss a few possible manifestations of inflated capitalization. First, one must recognize that the free cash flow is due to sales of products and services and is not directly affected by inflation of estate capitalization. Secondly, it is found from Eq. (1) to (4) that provided the capitalization $K$ and the value change rate $d\kappa/dt$ are affected similarly, the capital return rate is invariant, and does not trigger changes in management practices. Then, however, Eq. (6) is apparently violated. It must be complemented as

$$\int_{a}^{a+\tau} \frac{d\kappa}{dt} dt = \int_{a}^{a+\tau} \frac{dC}{dt} dt + \int_{a}^{a+\tau} \frac{dD}{dt} dt \qquad (7),$$

where $dD/dt$ refers to the rate of intangible market premium. The intangible market premium however can be liquidized only on the real estate market, not on the timber market. Unless the real estate market is exploited, the closed integral under periodic boundary conditions

$$\int_{a}^{a+\tau} \frac{dD}{dt} dt = 0 \qquad (8).$$

Further, the change rate of capitalization can be decomposed as

$$\frac{dK}{dt} = \frac{d\kappa}{dt} - \frac{dN}{dt} + \frac{dI}{dt} = \frac{dV}{dt} - \frac{dA}{dt} - \frac{dN}{dt} + \frac{dI}{dt} \qquad (9),$$

where $dN/dt$ is net cash flow, $dI/dt$ is the rate of capitalized investments, $dV/dt$ is the net value growth rate, and $dA/dt$ is the rate of amortizations.

Considering a scaling factor $(1+u)$ for the net value growth rate $dV/dt$ and combining it with the boundary condition (5) results as

$$\int_{a}^{a+\tau} \frac{dN}{dt} dt = (1+u) \int_{a}^{a+\tau} \frac{dV}{dt} dt - u \int_{a}^{a+\tau} \frac{dN}{dt} dt \qquad (10)$$

and

$$\int_{a}^{a+\tau} \frac{dC}{dt} dt = (1+u) \int_{a}^{a+\tau} \frac{dV}{dt} dt - u \int_{a}^{a+\tau} \frac{dN}{dt} dt - \int_{a}^{a+\tau} \frac{dA}{dt} dt$$

$$= (1+u) \int_{a}^{a+\tau} \frac{d\kappa}{dt} dt - u \int_{a}^{a+\tau} \frac{dC}{dt} dt \qquad (11)$$

It is found that as long as the capitalization premium $u$ is zero, Eq. (11) coincides with Eq. (6). Let us then consider the scaling factor $(1+u)$ for the net value growth rate

$$\langle r' \rangle = \frac{\int_a^{a+\tau} \frac{dC}{dt} dt}{\int_a^{a+\tau} K' dt} = \frac{(1+u)\int_a^{a+\tau} \frac{d\kappa}{dt} dt - u\left[\int_a^{a+\tau} \frac{dN}{dt} dt - \int_a^{a+\tau} \frac{dA}{dt} dt\right]}{(1+u)\int_a^{a+\tau} K dt}$$

$$= \frac{(1+u)\int_a^{a+\tau} \frac{d\kappa}{dt} dt - u\int_a^{a+\tau} \frac{dC}{dt} dt}{(1+u)\int_a^{a+\tau} K dt}$$

(12).

Provided the value growth sums up as free cash flow according to Eq. (6), Eq. (12) converts into

$$\langle r' \rangle = \frac{\int_a^{a+\tau} \frac{d\kappa}{dt} dt}{(1+u)\int_a^{a+\tau} K dt}$$

(13).

In other words, the inflated capitalization simply scales Eq. (4) by a factor of $1/(1+u)$.

However, there might be a possibility that Eq. (6) would become violated. This also would correspond to a violation of the periodic boundary condition (5): if the amount of harvesting does not sum up to the accumulated net growth, the capitalization is no longer periodic. Departure of periodicity makes that the integration can no longer be started from an arbitrary time as in Eq. (12). Instead, the value increment rate must be integrated from the establishment of a stand to maturity. In the absence of periodicity, Eq. (12) must be rewritten as

$$\langle r' \rangle = \frac{(1+u)\int_0^\tau \frac{d\kappa}{dt} dt - u\left[\int_0^\tau \frac{dN}{dt} dt - \int_0^\tau \frac{dA}{dt} dt\right]}{(1+u)\int_a^{a+\tau} K dt}$$

$$= \frac{(1+u)\int_0^\tau \frac{d\kappa}{dt} dt - u\int_0^\tau \frac{dC}{dt} dt}{(1+u)\int_0^\tau K dt}$$

(14).

If the net cash flow approaches zero in the absence of harvesting, Eq. (14) turns into

$$\langle r' \rangle = \frac{\int_0^\tau \frac{d\kappa}{dt} dt + \frac{u}{1+u}\int_0^\tau \frac{dA}{dt} dt}{\int_0^\tau K dt}$$

(15).

Eq. (15) shows that there is no linear scaling in the capital return rate, in relation to Eq. (4). Instead, there is a slight increment in capital return rate since the goodwill in capitalization applies to the net growth rate but not to amortizations.

It is found from Eq. (14) that the creation of free cash flow deteriorates the return rate of capital. In other words, it reduces the capitalization premium in the numerator of Eq. (14), resulting ultimately in the linear scaling appearing in Eq. (13). Harvesting deteriorates value, but if harvesting does not sum up to the accumulated net growth, the periodic boundary conditions are violated. This phenomenon is here denoted as a *continuity problem* of value creation.

Two different forestry strategies are here applied. Firstly, only the timber market is entered, and consequently, the deterioration of goodwill value along with harvesting according to Eq. (8) is accepted. Consequently, the capital return rate is given by Eq. (13). This strategy is here denoted as the "Timber Sales" – strategy (TS). Secondly, the real estate market is entered at stand maturity. This corresponds to abandoning the periodic boundary conditions (5) and (6) within the financial perspective of the operating agent. Then, also Eq. (8) would become abandoned. Correspondingly, Eq. (14) can be applied without reference to Eqs. (5) and (6), but it does not necessarily lead to Eq. (15) – goodwill deterioration along with eventual thinnings is accepted if necessary for maximizing the total return. This strategy is here denoted as the "Real Estate" – strategy (RE). Again, depending on whether harvesting is excluded in Eq. (14), the outcome may or may not approach Eq. (15). Importantly, the maturity age possibly can be controlled through thinnings, at the expense of a capital return rate deficiency.

*2.2. The two datasets applied*

Two different sets of initial conditions have been described in four earlier investigations, together forming 16 different sets of initial conditions [45,42,43,32]. Firstly, a group of nine setups was created, containing three tree species and three initial sapling densities [43]. The idea was to apply the inventory-based growth model as early in stand development as it is applicable, to avoid approximations of stand development not grounded on the inventory-based growth model [46]. This approach also allowed an investigation of a wide range of stand densities, as well as a comprehensive description of the application of three tree species. The exact initial conditions here equal the ones recommended in [43], appearing there in Figures 8 and 9.

Secondly, seven wooded, commercially unthinned stands in Vihtari, Eastern Finland, were observed at the age of 30 to 45 years. The total stem count varied from 1655 to 2451 per hectare. A visual quality approximation was implemented. The number of stems deemed suitable for growing further varied from 1050 to 1687 per hectare. The basal area of the acceptable-quality trees varied from 28 to 40 m$^2$/ha, in all cases dominated by spruce (*Picea abies*) trees.

The two strategies discussed above are applied to both datasets. A proportional goodwill $(1+u) = (1+1/2)$ is applied according to Eqs. (13) and (14). The inflation factor is somewhat arbitrary, but it is based on recent observations [5,7,9], including very recent observations by the author: large, productive forest estates appear to change owners at 150% of the fair forestry value determined by professionals. In addition to the two market strategies, eventual thinning restrictions are discussed.

3. Results

Figs. 1, 2, and 3 show the expected value of the capital return rate within stands of three tree species where the growth model is applied as early as applicable. In any of the three tree species, commercial thinnings do not enter into the Real Estate – strategy. Correspondingly Eq. (14) coincides with Eq. (15), and thinning restrictions are irrelevant. In the absence of thinning restrictions,

stands to be harvested according to Timber Sales - strategy (Eq. (13)) do enter thinnings, with one exception. It is found that the achievable capital return rate is in the order of 50% greater in the Real-estate strategy, corresponding to shorter rotation ages.

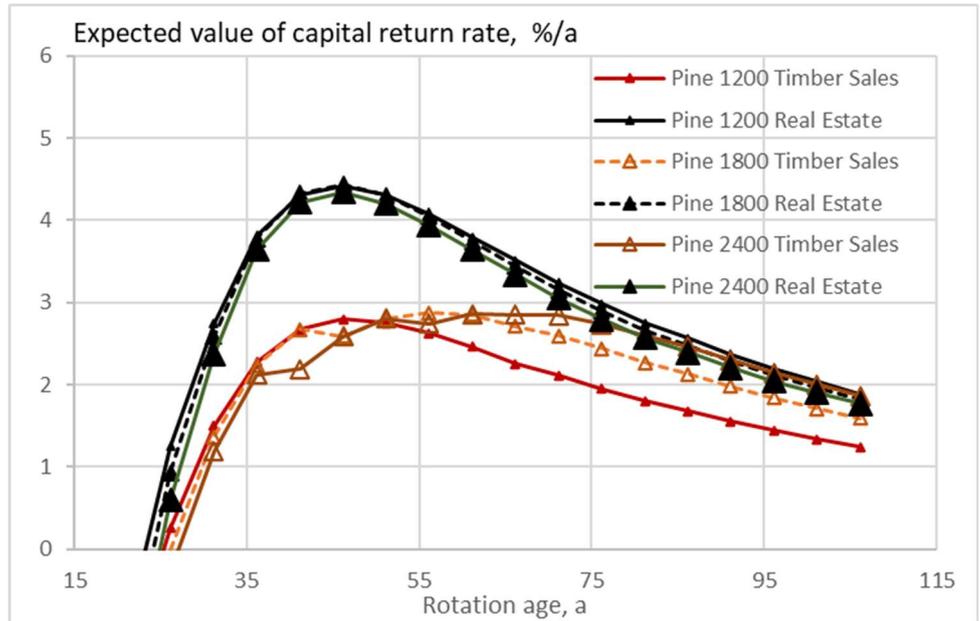

**Figure 1.** The expected value of capital return rate on pine (*Pinus sylvestris*) stands of different initial sapling densities, as a function of rotation age, when the growth model is applied as early as applicable, along with the Real Estate – strategy (Eq. (14)), and the Timber Sales – strategy (Eq. (13)).

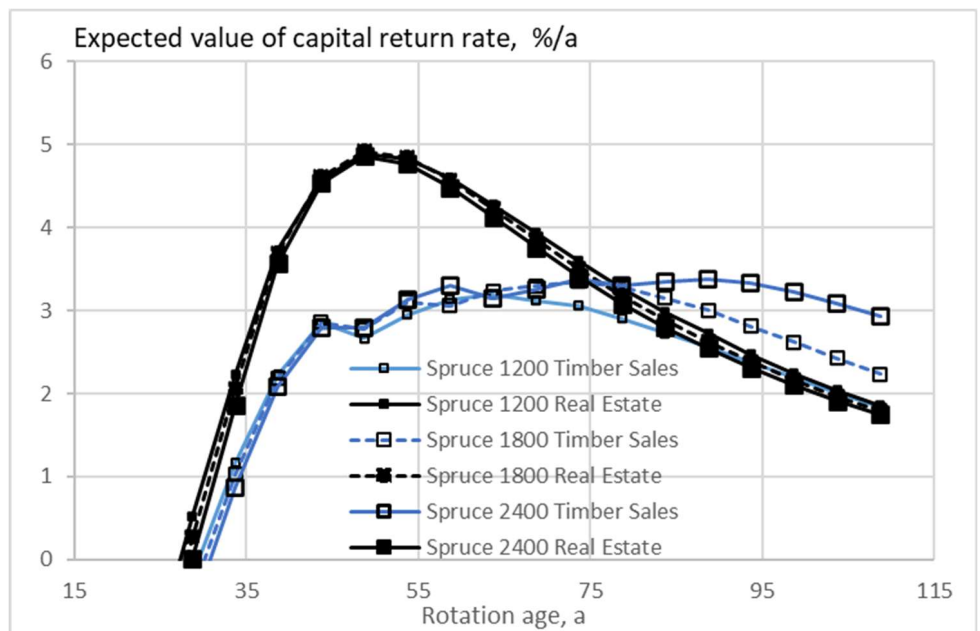

**Figure 2.** The expected value of capital return rate on spruce (*Picea Abies*) stands of different initial sapling densities, as a function of rotation age, when the growth model is applied as early as applicable, along with the Real Estate – strategy (Eq. (14)), and the Timber Sales – strategy (Eq. (13)).

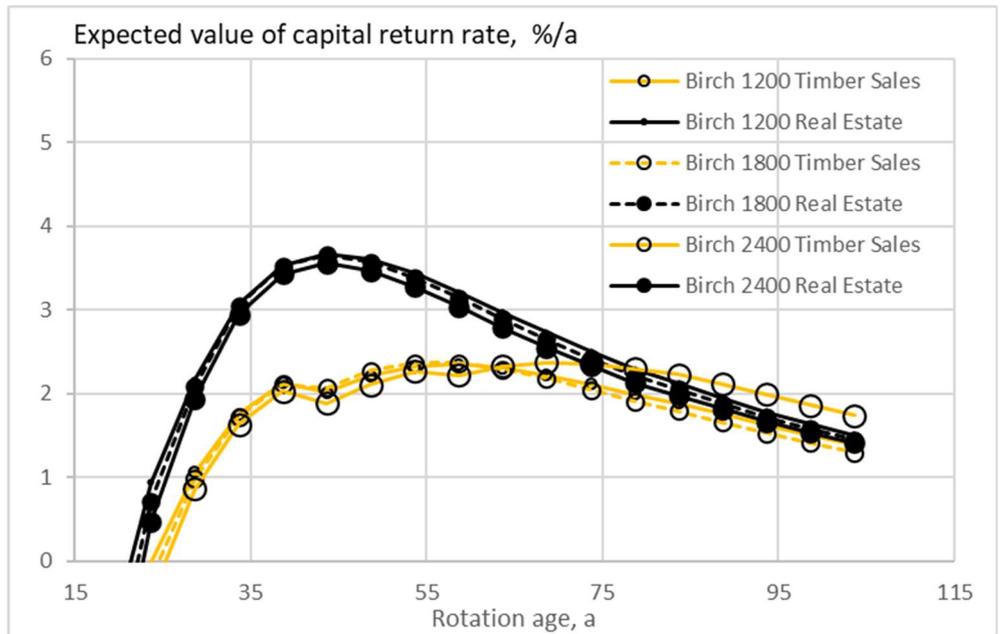

**Figure 3.** The expected value of capital return rate on birch (*Betula pendula*) stands of different initial sapling densities, as a function of rotation age, when the growth model is applied as early as applicable, along with the Real Estate – strategy (Eq. (14)), and the Timber Sales – strategy (Eq. (13)).

Figs. 4 and 5 show the expected value of the capital return rate within seven stands first observed at the age of 30 to 45 years, in the presence of inflated capitalization and eventual thinning restrictions. The capital return rate according to Eq. (14) within stands prepared for sale (Real Estate – strategy), Fig. 4) is consistently greater than that for stands to be harvested (Timber Sales – strategy, Fig. 5) according to Eq. (13). No commercial thinnings are entered according to the RE strategy. Correspondingly, Eq. (14) coincides with Eq. (15), and thinning restrictions are irrelevant. In the absence of thinning restrictions, stands to be harvested according to TS strategy (Eq. (13)) always enter thinnings. The rotation times according to Eq. (15) are consistently shorter, and capital return rates in the order of 50% greater.

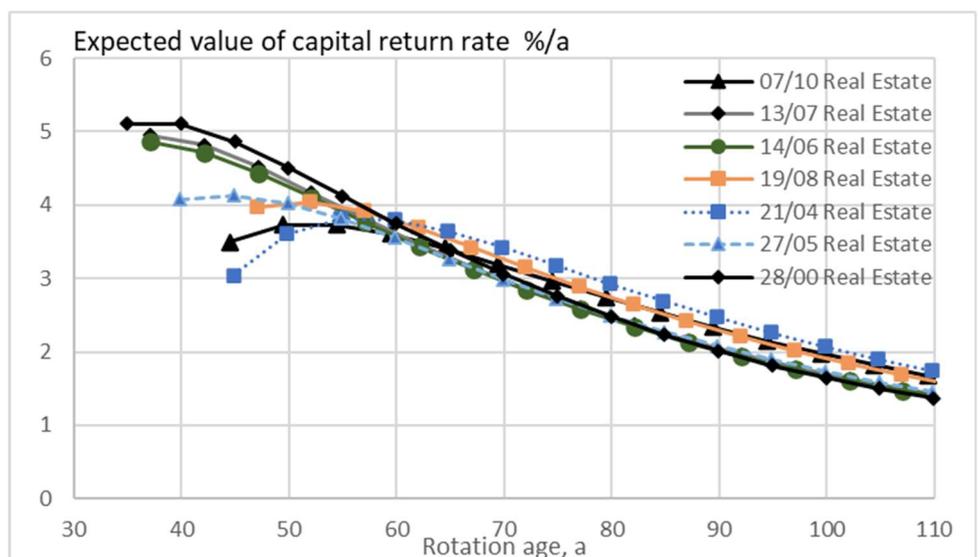

**Figure 4.** The expected value of capital return rate, as a function of rotation age,

when the growth model is applied to seven observed wooded stands, along with the Real Estate – strategy (Eq. (14)). Thinnings do not enter, and Eq. (14) coincides with Eq. (15). The numbers in legends identify stands and observation plots.

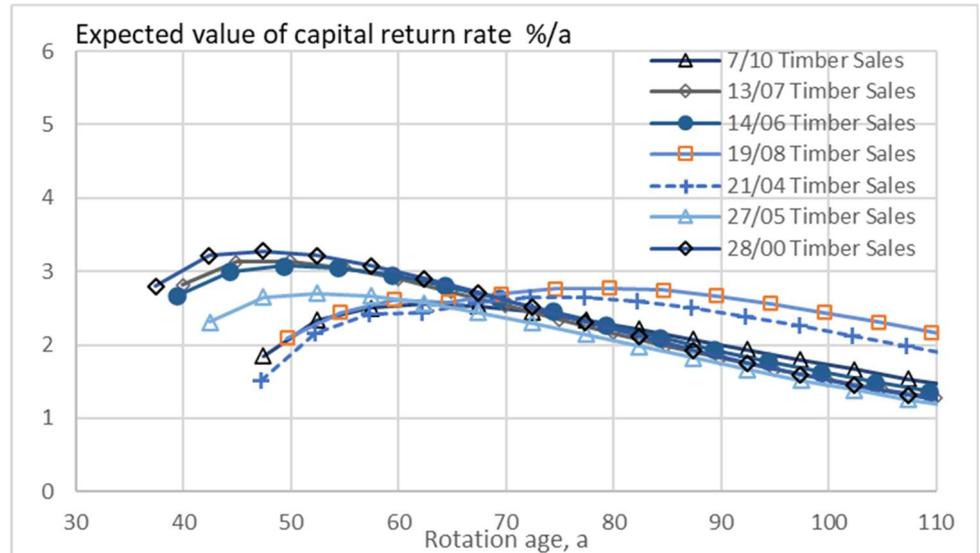

**Figure 5.** The expected value of capital return rate, as a function of rotation age, when the growth model is applied to seven observed wooded stands, along with the Timber Sales – strategy (Eq. (13)). Thinnings do enter in all cases. The numbers in legends identify stands and observation plots.

## 4. Discussion

Two strategies for boreal forestry have been discussed at stand level. The Timber Sales – strategy can be naturally applied at stand level. The Real Estate – strategy however should be applied at the estate level. Correspondingly, the stand-level treatment is somewhat problematical and would apply as such only in even-aged estates. In the case of uneven-aged estates, robustness against varying rotation ages at the stand level would be beneficial.

Another issue is that the age structure of an estate is not necessarily the only factor contributing to the timing of entering the real estate market – a major timing contributor may be family reasons, possibly related to the transfer of property to the next generation.

Let us discuss the robustness of the Real Estate - strategy regarding rotation ages. It might be of interest to clarify the possibility of extending the rotation times within the RE strategy. It would allow older stands within an estate to hold while younger stands mature. On the other hand, such a possibility would create possibilities for selecting a suitable time for entering the real estate market, not necessarily directly due to estate age structure.

A possibility for extending rotation ages within the RE strategy might be the application of commercial thinnings. Then, Eq. (14) would no longer coincide with Eq. (15) as the net cash flow does not sum up to zero. In Figs. 1 to 4, the suitable rotation ages vary from 40 to 60 years, without thinnings. It is of interest how the rotation ages possibly could be extended by 20 years without any major loss of the expected value of capital return rate on the stand level.

It is found from Figs. 6 to 8 that in stands where the growth model is applied as applicable, thinnings from above can extend rotations with a minor loss in the financial return, and there is a significant difference of this modified RE strategy to the corresponding outcome of the TS strategy. The same partially applies to stands first observed at age 30 to 45 years in Fig. 9. However, in the case of fertile stands reaching the maximum financial return in Fig. 4 soon after the time of stand observation, a greater loss results from the extension of the rotation time is found in Fig. 9. This possibly indicates that from the viewpoint of extending the rotation time, these stands are, at the time of observation, overdue for the best timing of thinning.

While the results of this paper indicate that, in the presence of proportional goodwill in estate prices, large-scale harvesting becomes non-profitable within the Real Estate – strategy, a question arises regarding the necessity of climate change mitigation arrangements. The answer appears to be two-fold. Eventual carbon rent would change the thinning practices appearing in Figs. 6 to 9. On the other hand, timberland end users, like insurance companies and investment trusts, probably will retain the Timber Sales – strategy. Correspondingly, the operations of such institutions remain as subjects for mitigation programs [47,32,33,41].

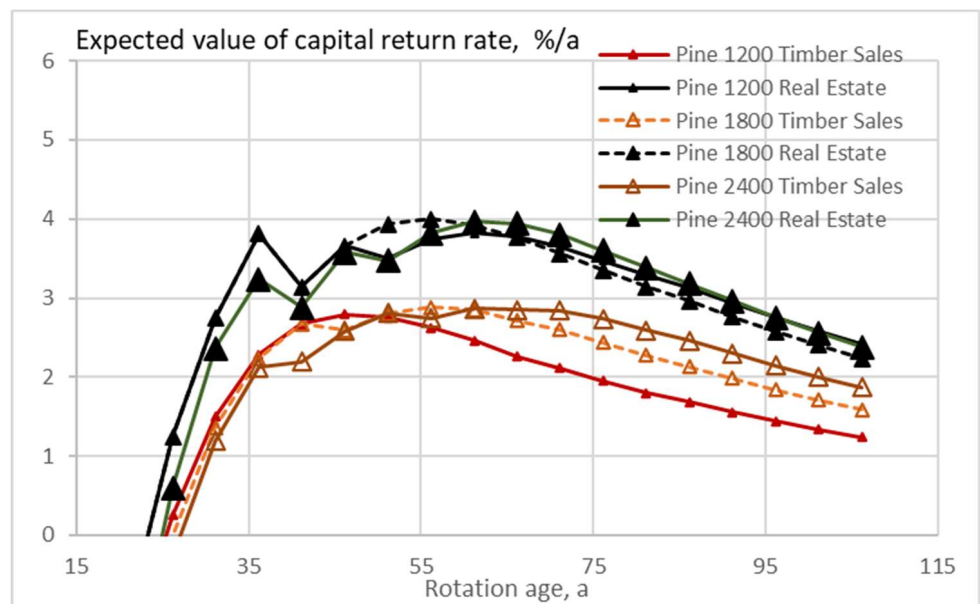

**Figure 6.** The expected value of capital return rate on pine (*Pinus sylvestris*) stands of different initial sapling densities, as a function of rotation age, when the growth model is applied as early as applicable, along with the Real Estate – strategy (Eq. (14)), and the Timber Sales – strategy (Eq. (13)). Thinnings have been introduced into the RE strategy to extend the feasible rotations by 20 years.

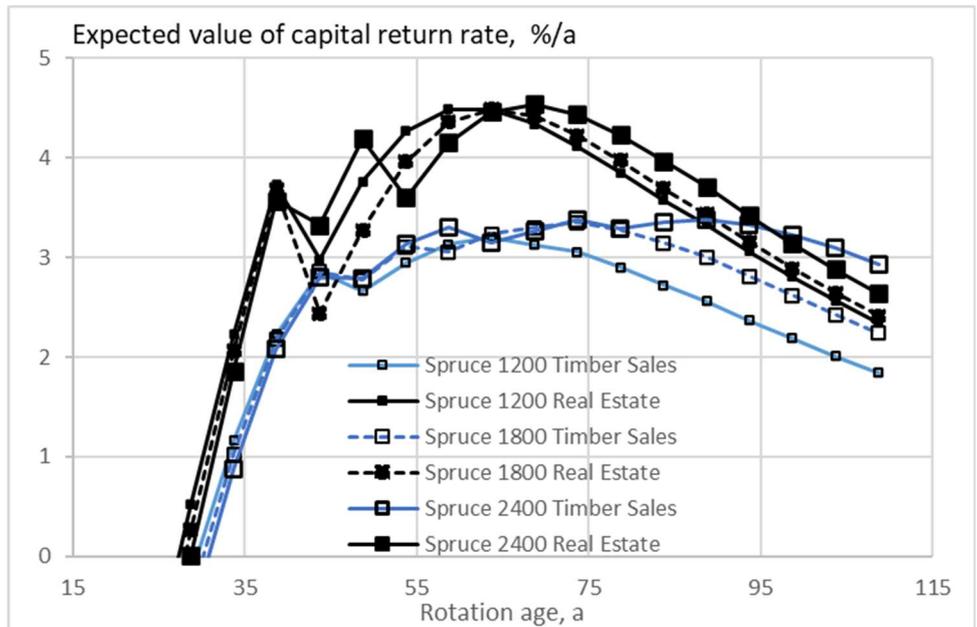

**Figure 7.** The expected value of capital return rate on spruce (*Picea Abies*) stands of different initial sapling densities, as a function of rotation age, when the growth model is applied as early as applicable, along with the Real Estate – strategy (Eq. (14)), and the Timber Sales – strategy (Eq. (13)). Thinnings have been introduced into the RE strategy to extend the feasible rotations by 20 years.

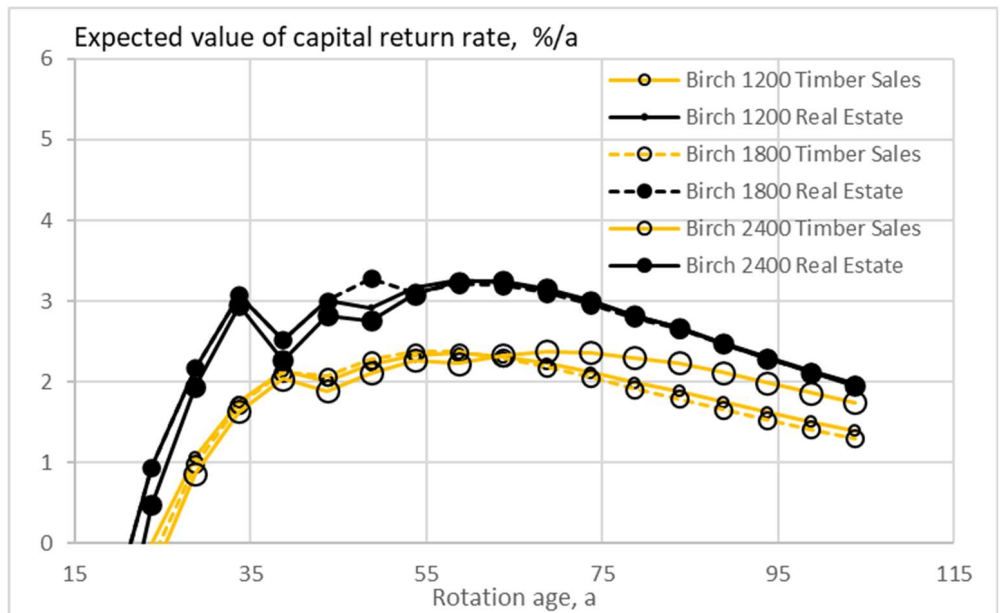

**Figure 8.** The expected value of capital return rate on birch (*Betula pendula*) stands of different initial sapling densities, as a function of rotation age, when the growth model is applied as early as applicable, along with the Real Estate – strategy (Eq. (14)), and the Timber Sales – strategy (Eq. (13)). Thinnings have been introduced into the RE strategy to extend the feasible rotations by 20 years.

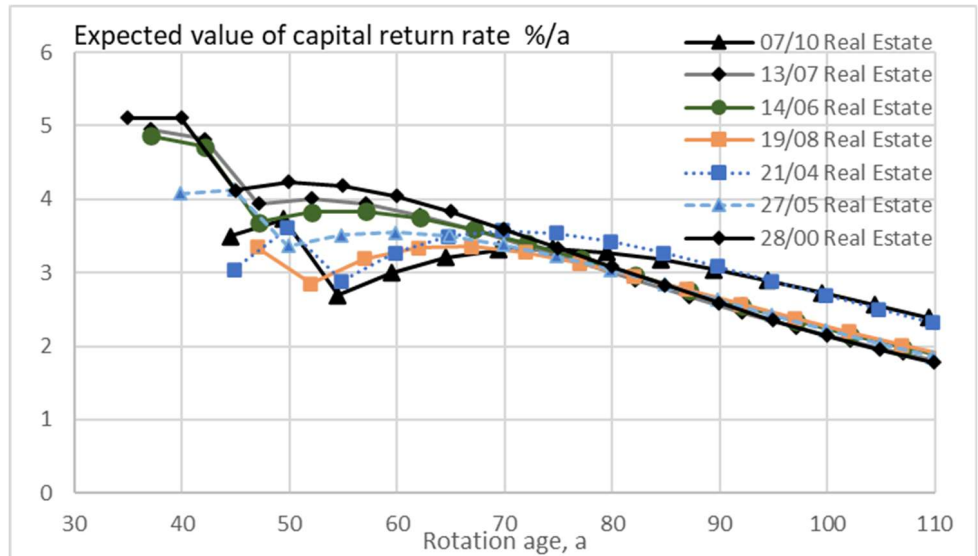

**Figure 9.** The expected value of capital return rate, as a function of rotation age, when the growth model is applied to seven observed wooded stands, along with the Real Estate – strategy (Eq. (14)). The numbers in legends identify stands and observation plots. Thinnings have been introduced into the RE strategy to extend the feasible rotations by 20 years.

## 5. Conclusions

Two strategies for boreal forestry with goodwill in estate capitalization were introduced. The strategy focusing on Real Estate (RE) was financially superior to Timber Sales (TS). The feasibility of the RE requires the presence of forest land end users in the real estate market, like insurance companies or investment trusts, and the periodic boundary condition does not apply. Commercial thinnings did not enter the RE strategy in a stand-level discussion. However, they may appear in estates with a variable age structure and enable an extension of stand rotation times.

**Funding:** This research was partially funded by Niemi Foundation.

**Data Availability Statement:** Datasets used have been introduced in earlier papers referenced above.

**Conflicts of Interest:** The author declares no conflict of interest.